\documentclass[prl,graphicx,twocolumn,showpacs]{revtex4-1}

\usepackage{amsmath}
\usepackage{enumerate}
\usepackage{graphicx}
\usepackage{mathbbol}
\usepackage{amsfonts}
\usepackage{natbib}

\usepackage{bm}

\usepackage{color}

\draft % marks overfull lines with a black rule on the right

\begin{document}

\title{Universal Order Statistics of Random Walks}

\author{Gr\'egory Schehr}
\email[]{gregory.schehr@th.u-psud.fr}
\affiliation{Laboratoire de Physique Th\'eorique d'Orsay, Universit\'e Paris Sud 11 and CNRS}

\author{Satya N. Majumdar}
\email[]{majumdar@lptms.u-psud.fr}
\affiliation{Univ. Paris-Sud, CNRS, LPTMS, 91405 Orsay Cedex, France}

\date{\today}

\begin{abstract}

We study analytically the order statistics of a time series generated by the
successive positions of a symmetric random walk of $n$ steps with
step lengths of finite variance $\sigma^2$. We
show that the statistics of the gap 
$d_{k,n}=M_{k,n} -M_{k+1,n}$ between  the $k$-th and the $(k+1)$-th maximum
of the time series becomes {\em stationary}, i.e, independent of $n$ as $n\to 
\infty$ and exhibits a rich, universal behavior.
The mean stationary gap (in units of $\sigma$) exhibits a universal algebraic 
decay
for large $k$, $\langle d_{k,\infty}\rangle/\sigma\sim 
1/\sqrt{2\pi k}$, 
independent of the details of the jump distribution.
Moreover, the probability density (pdf) of 
the stationary gap exhibits scaling, 
${\rm Pr}(d_{k,\infty}=\delta)\simeq (\sqrt{k}/\sigma) P(\delta 
\sqrt{k}/\sigma)$, in the scaling regime when $\delta\sim \langle 
d_{k,\infty}\rangle\simeq \sigma/
\sqrt{2\pi k}$.
The scaling function $P(x)$ is universal and has an  
unexpected power law
tail, $P(x) \sim x^{-4}$ for large $x$. For $\delta \gg \langle 
d_{k,\infty}\rangle$ the scaling breaks down and the pdf gets
cut-off in a nonuniversal way. Consequently, the moments
of the gap exhibit an unusual multi-scaling behavior.

\end{abstract}


\maketitle

During the last fifty years, extreme value statistics (EVS), the statistics of 
the maximum or the minimum of a set of random variables, have found many 
applications, ranging from engineering~\cite{gumbel} to environmental 
sciences~\cite{katz} or finance~\cite{embrecht, bouchaud_satya}, where rare 
and extreme events may have drastic consequences. It was  
demonstrated \cite{bm97} that EVS also plays a major role in the physics of 
complex 
and disordered systems. Therefore finding the distribution of the maximum 
$x_{\max}$ (or the minimum $x_{\min}$) of a set of $n+1$ random variables 
$\{x_0, x_1, x_2, \cdots, x_n\}$ has been the subject of intense activity not 
just 
for independent and identically distributed (iid) random 
variables~\cite{gumbel}, but also recently for {\em strongly correlated} 
random 
variables~\cite{dean_majumdar, pldmonthus, pld_carpentier, satya_airy, 
schehr_airy, gyorgyi, satya_yor,comtet_precise,schehr_rsrg} that are often more 
relevant  
in physical contexts.

While the statistics of the extremum $x_{\rm max}$ (or $x_{\rm 
min}$) is important, they concern the fluctuations of a single value among a 
typically large sample and a natural question is then: are these extremal 
values isolated, i.e., far away from the others, or are there many other 
events close to them? Such questions have led to the study of the density of 
states of near-extreme events \cite{sanjib_nearextreme, 
burkhardt_nearextreme}. This is, for instance, a crucial question  
in disordered 
systems, where the low temperature properties are governed by excited 
states close to the ground state.  A natural 
way to characterize this phenomenon of crowding of near-extreme 
events is via the order statistics, i.e., arranging the
random variables $x_m$'s in decreasing order of magnitude  
$M_{1,n} > ... > M_{k,n} > 
... > M_{n+1,n}$ where $M_{k,n}$ denotes the $k$-th maximum 
of the set $\{x_0,x_1,\dots,x_n\}$. Evidently, $x_{\rm max} = M_{1,n}$, while 
$x_{\rm min} = M_{n+1,n}$. A set of useful observables that are naturally 
sensitive to the crowding of extremum are the gaps between 
the consecutive ordered maxima: $d_{k,n} = M_{k,n} - M_{k+1,n}$ denoting
the $k$-th gap.
 
While the study of order (or gap) statistics has received considerable 
interest in statistics literature, e.g., in the context of 
system reliability~\cite{order_book}, the available results are 
restricted only to iid variables. In contrast, there hardly exist 
analytical results for the gap statistics for {\em strongly correlated} 
random variables. The importance of order statistics for
such correlated variables 
came up recently in several physical contexts, notably 
in the study of the branching Brownian motion~\cite{derrida_bbm}
and also for $1/f^\alpha$ signals~\cite{racz_order} with
an application to the statistical analysis of cosmological
observations \cite{tremaine_cosmo}. Any solvable model
for the order statistics for correlated variables would
thus be welcome and this Letter takes a step in that direction.

\begin{figure}[ht]
%\begin{center}
\includegraphics[width = 0.5\linewidth]{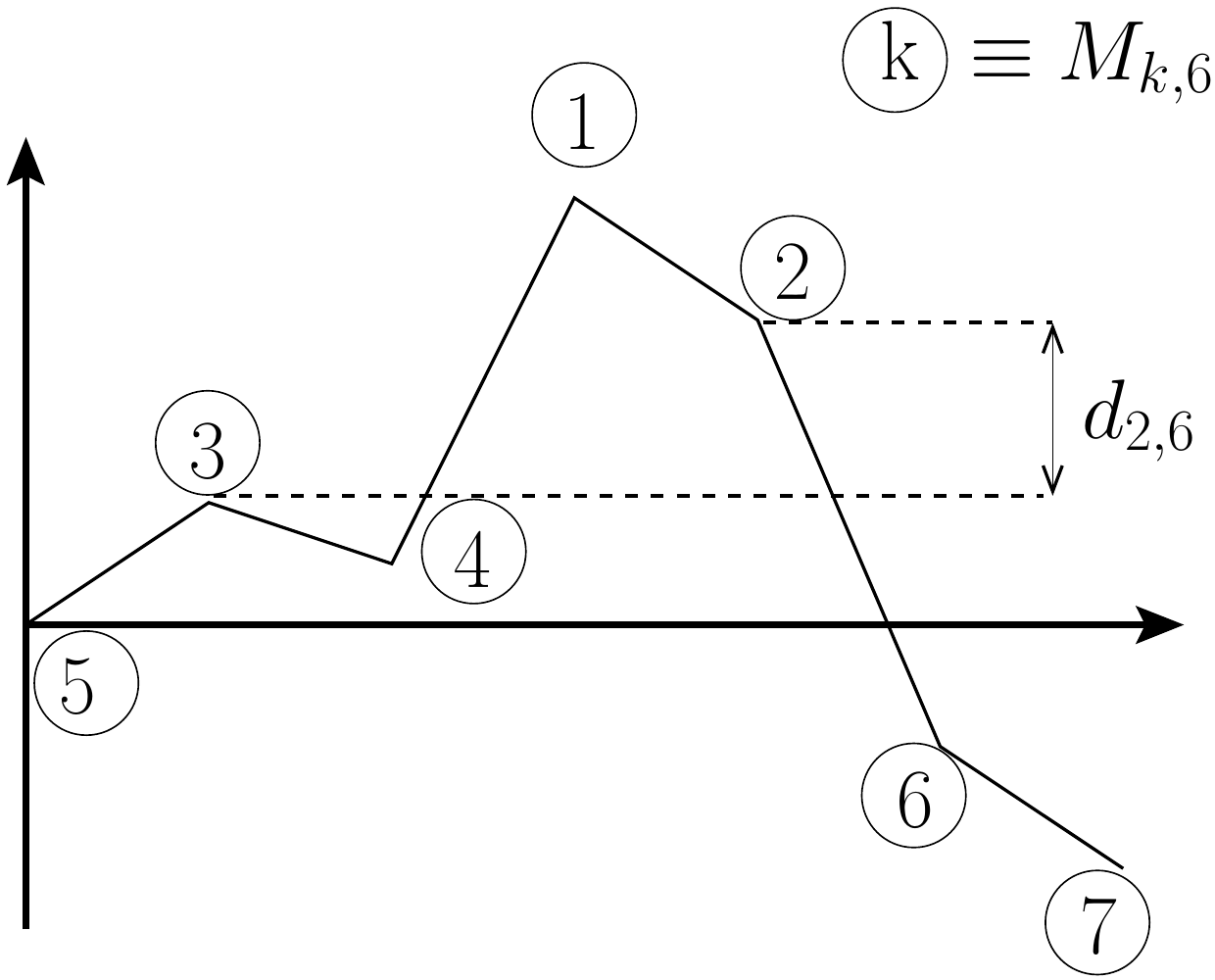}
\caption{A realization of a random walk of $n=6$ steps. 
We denote by $M_{k,6}$ the $k$-th maximum and focus in 
particular on the gaps $d_{k,n} = M_{k,n} - M_{k+1,n}$. Note that $x_0$ is taken into account in the statistics.}\label{cartoon_rw}
%\end{center}
\end{figure}

In this Letter, we present exact analytical results for the order statistics 
and the gap distribution of a time series $\{x_0,x_1,\ldots,x_n\}$ 
where $x_m$ represents the position of a random walker at discrete time $m$.
The walker starts at $x_0=0$ at time $0$ and at each discrete step evolves via 
$x_m= x_{m-1} +\eta_m$, where the noise $\eta_m$'s are iid 
jump lengths each drawn from a symmetric and continuous distribution $f(\eta)$
with zero mean and a finite variance 
$\sigma^2=\int_{-\infty}^{\infty} \eta^2\, f(\eta)\, d\eta$.
Even though the jump lengths are uncorrelated, the entries
$x_m$'s are clearly correlated and represent perhaps
the simplest, yet most ubiquitous correlated time series (discrete-time
Brownian motion)
with a large variety of applications~\cite{feller_book, spitzer_book}, 
including for instance in queuing theory~\cite{queuing} -- where 
$x_m$ represents the length of a single server queue at time $m$ -- or in 
finance 
where $x_m$ represents the logarithm of the price of a stock 
at time $m$~\cite{finance}. Even for this relatively simple correlated
time series, we show that the gap distribution exhibits a rather
rich and universal behavior. 

It is useful to summarize our main results. For large $n$, one finds that 
$\langle M_{k,n}\rangle/\sigma = \sqrt{2n/\pi} + {\cal O}(1)$, independent of 
$k$. Thus the property of the crowding of extremum ($k$-dependence) 
is not captured by the statistics of the maxima $M_{k,n}$ themselves,
at least to leading order for large $n$. The simplest observable
that is sensitive to the crowding phenomenon is the gap,
$d_{k,n} = M_{k,n} -
M_{k+1,n}$ (see Fig.~\ref{cartoon_rw}). Our main result is to show that the
statistics of the scaled gap $d_{k,n}/\sigma$ becomes stationary, 
i.e., independent of $n$ for 
large $n$, but
retains a rich, nontrivial $k$ dependence which becomes {\em 
universal} for large $k$, i.e. independent of the details of the jump 
distribution
$f(\eta)$.
We compute the stationary mean gap $\bar d_k=\langle d_{k,\infty}\rangle$ 
exactly for all 
$k$, for arbitrary $f(\eta)$ and show that, when
expressed in units of $\sigma$, it has a universal algebraic tail, $\bar 
d_k/\sigma\approx 
1/\sqrt{2\pi k}$ for large $k$. Next, we compute exactly the full
pdf of the stationary gap $p_k(\delta)={\rm Pr}(d_{k,\infty}=\delta)$
for the exponential jump distribution, 
$f(\eta)=b^{-1}\exp\left(-|\eta|/b\right)$
and show that for large $k$, there is a scaling regime when 
$\delta\sim \langle
d_{k,\infty}\rangle\simeq \sigma/
\sqrt{2\pi k}$ where the pdf scales as, $p_k(\delta)\simeq 
(\sqrt{k}/\sigma) P(\delta
\sqrt{k}/\sigma)$, with a nontrivial scaling function
\begin{equation}\label{exact_F} 
P(x) = 4\big[\sqrt{\frac{2}{\pi}}(1+2x^2) -
e^{2x^2}x(4x^2+3) {\rm erfc}(\sqrt{2}x)\big] 
\,, 
\end{equation} 
where ${\rm
erfc}(z) = (2/\sqrt{\pi})\int_z^\infty e^{-t^2} \, dt$ is the complementary   
error function. While we were unable to compute the gap pdf for
arbitrary $f(\eta)$, our numerical simulations provide
strong evidence that the scaling function $P(x)$ in Eq. (\ref{exact_F})
is actually universal, i.e., independent of $f(\eta)$. 
Somewhat unexpectedly, we find that this universal scaling function 
has an algebraic tail $P(x)\sim x^{-4}$ for large $x$.
For $\delta \gg \langle
d_{k,\infty}\rangle\simeq \sigma/
\sqrt{2\pi k}$, the pdf gets cut-off in a nonuniversal fashion.
This is shown to have interesting consequences for the
moments of the stationary gap:
$\langle d_{k,n}^p \rangle \sim 
k^{-\frac{p}{2}}$ for $p<3$,
while $\langle d_{k,n}^p \rangle \sim k^{-\frac{3}{2}}$ for $p>3$.

We start with the statistics of the $k$-th 
maximum $M_{k,n}$ of the random walk $x_m=x_{m-1}+\eta_m$ 
of $n$ steps, starting from the initial value $x_0=0$.
The goal is to write down an evolution equation for
the cumulative distribution
of the $k$-th maximum $F_{k,n}(x) = {\rm Pr}[M_{k,n} \leq x]$.
The event $M_{k,n} \leq x$ means that we have at most
$(k-1)$ points above the level $x$ between step $1$ and $n$.
To keep track of this event, it is convenient first to define an
auxiliary quantity $q_{k,n}(x)$ 
denoting the probability that 
the random walk, starting at $x_0=x$, has $k$
points below $0$ from step $1$ to step $n$. 
It is then easy to see that $F_{k,n}(x)$ can be expressed as the sum
\begin{eqnarray}\label{rel_FQ}
 F_{k,n}(x) = 
 \begin{cases}
&\, \sum_{m=0}^{k-1} q_{m,n}(x) \;,\hspace*{1.05cm} x  > 0 \\
&\, \sum_{m=0}^{k-2} q_{n-m,n}(-x) \;,\; x < 0   \;.,
\end{cases}
 \end{eqnarray}
where we used that $f(\eta)$ is symmetric and continuous.

\begin{figure}
\begin{center}
\includegraphics[width = 1. \linewidth]{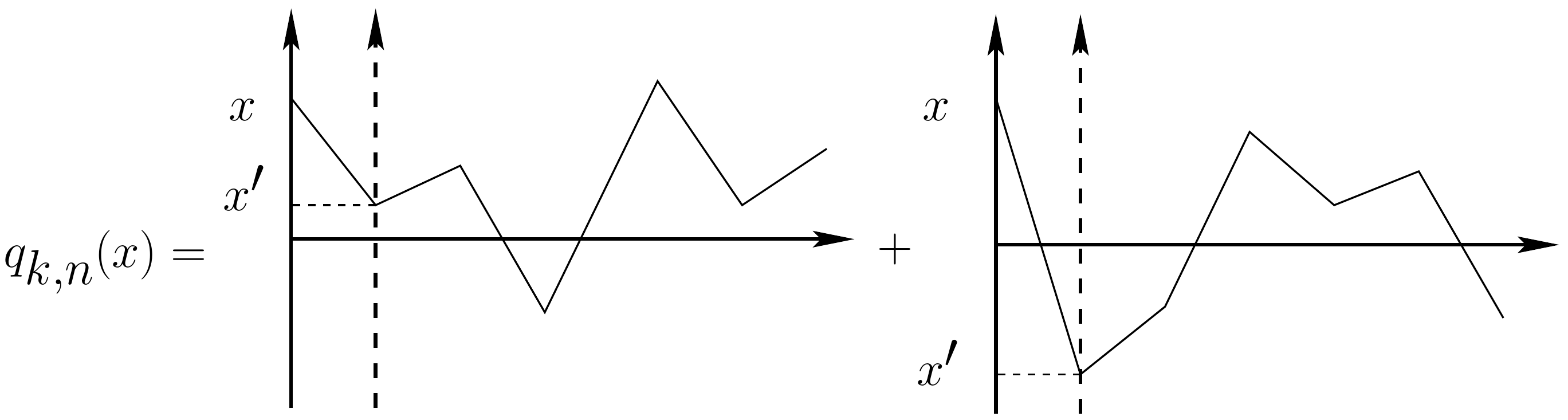}
\caption{Illustration of the backward equation in 
Eq. (\ref{back_fp1}).}\label{cartoon_backward}
\end{center}
\end{figure}

The next step is to write a backward recurrence equation for 
$q_{k,n}(x)$ by considering the stochastic jump $x\to x'$ at the first step (see
Fig. \ref{cartoon_backward})
and then subsequently using the Markov property of the evolution.
One gets, for $n\ge 1$,
\begin{eqnarray}\label{back_fp1}
q_{k,n}(x) &=& \int_0^\infty q_{k,n-1}(x') f(x'-x) \, dx' \nonumber \\
&+& \int_{-\infty}^0 q_{n-k,n-1}(-x') f(x'-x) dx'  \;,
\end{eqnarray}
starting from $q_{0,0}(x) = 1$. The first term corresponds to a jump from
$x>0$ to $x'>0$ while the second term corresponds to a jump from
$x>0$ to $x'<0$ (see Fig. \ref{cartoon_backward}).

The integral equation (\ref{back_fp1}) is of the Wiener-Hopf type
which are generically hard to solve for arbitrary jump distribution
$f(x)$. However, for the special case 
$f(x) = \frac{1}{2 b} \exp{(-|x|/b)}$ (with 
$b = \sigma/\sqrt{2}$), using the 
useful property, $f''(x)=b^2 f(x)- b^2 \delta(x)$, 
we were able to reduce this integral
equation into a differential recurrence equation
which can subsequently be solved by generating function method.
Skipping details~\cite{supp_mat}, we get
\begin{eqnarray}
&&\tilde q(z,s,x) = \sum_{n=0}^\infty \sum_{k=0}^n s^n z^k q_{k,n}(x) = 
\frac{1}{1-s} \\
&& +\left(\frac{1}{\sqrt{(1-s)(1-zs)}} - \frac{1}{1-s} \right)
\exp{\left(-\sqrt{2(1-s)} \frac{x}{\sigma}\right)}  \;, \nonumber
\end{eqnarray}
from which, using Eq. (\ref{rel_FQ}), the 
$p$-th moment of the $k$-th maximum $\langle M_{n,k}^p\rangle$ 
can be extracted. In particular for $p=1$ we get~\cite{supp_mat} 
\begin{eqnarray}\label{exact_mkn}
\frac{\langle M_{k,n} \rangle}{\sigma} = \sum_{m=k}^{n-k+1} 
\frac{\Gamma\left(m + \frac{1}{2} \right)}{ \sqrt{2\pi} m !} \;.
\end{eqnarray}
It follows from (\ref{exact_mkn}) that for large $n$, ${\langle 
M_{k,n} \rangle}/{\sigma} \sim \sqrt{2\,n/\pi}$, independently of $k$, while 
the average gap $\langle d_{k,n}\rangle$ is given by
\begin{eqnarray}\label{exact_mean_gap_exp}
\frac{\langle d_{k,n} \rangle}{\sigma} = \left(\frac{\Gamma\left(k + \frac{1}{2} 
\right)}{ \sqrt{2\pi} k !} + 
\frac{\Gamma\left(n-k + \frac{3}{2} \right)}{ \sqrt{2\pi} (n-k+1) !} \right)\; 
.
\end{eqnarray}
Note that 
$\langle d_{k,n} \rangle = \langle d_{n-k+1,n} \rangle$ reflecting
the up-down (max-min) symmetry of the walk. 
Interestingly, as $n~\to~\infty$,  
$\langle d_{k,n} \rangle$ approaches a finite 
value 
\begin{eqnarray}\label{dtilde_exp}
\lim_{n \to \infty} \frac{\langle d_{k,n} \rangle}{\sigma} = 
\frac{\Gamma\left(k + \frac{1}{2} \right)}{ \sqrt{2\pi} k !}\; . 
\end{eqnarray}
In addition, for large $k$, 
\begin{eqnarray}\label{dtilde_asympt}
\lim_{n \to \infty} \frac{\langle d_{k,n} \rangle}{\sigma} = 
\frac{1}{\sqrt{2 \pi k}} + {\cal O}(k^{-1})\;.
\end{eqnarray}
Next we show that the result (\ref{dtilde_asympt}) is actually universal
and holds for arbitrary symmetric and continuous jump distribution $f(x)$.
To make progress for general $f(x)$, we came across a very useful
combinatorial identity known as Pollaczek-Wendel identity~\cite{pollaczek, wendel}.
Using this identity and a few manipulations~\cite{supp_mat}, we were able to
derive the following exact result
\begin{eqnarray}\label{exact_mean_gap}
&&\lim_{n \to \infty} \langle d_{k,n} \rangle =\bar d(k)  = 
\frac{\sigma}{\sqrt{2 \pi}} \frac{\Gamma(k+\frac{1}{2})}{\Gamma(k+1)}  \\
&& - \frac{1}{\pi k} \int_0^\infty \frac{dq}{q^2} 
\left[ [\hat f (q) ]^k - \frac{1}{(1 + \frac{\sigma^2}{2}q^2)^k} \right] 
\label{exact_gap} \;, \nonumber
\end{eqnarray} 
where $\hat f(q) =
\int_{-\infty}^\infty e^{iq\eta} f(\eta) \, d\eta$ is the Fourier transform
of the jump distribution. 
The asymptotic analysis of 
(\ref{exact_mean_gap}) gives the universal result~\cite{supp_mat}
\begin{eqnarray}\label{first_moment_universal}
{\bar d(k)}/{\sigma} \sim {\left({2 \pi k}\right)}^{-\frac{1}{2}} \;, \; k \gg 1 \;,
\end{eqnarray}
independent of $f(\eta)$. This $k^{-\frac{1}{2}}$ dependence of $\bar d_k$ (\ref{first_moment_universal}) was actually noticed
in the numerical study of periodic random walks in Ref. \cite{racz_order} and was also conjectured to be exact, based on scaling arguments.

This result naturally raises the question whether only the first moment 
of the
gap is universal, or perhaps the universality extends even to the pdf
of the gap, once it is scaled by the nonuniversal
scale factor $\sigma$. This led us next to investigate 
the full pdf of $d_{k,n}$. It is convenient first to
consider the joint cumulative distribution 
$S_{k,n}(x,y) = {\rm
Pr}[M_{k,n} > y, M_{k+1,n} <
x]$, with $y > x$. If we can compute this, then the
gap pdf $P_{k,n}(d_{k,n}=\delta)$ can be obtained 
from the relation 
\begin{equation}\label{gappdf}
{P}_{k,n}(\delta) =  - \int_{\mathbb{R}^2}
\frac{\partial^2 {S}_{k,n}(x,y)}{\partial x \partial y}
\theta(y-x) \delta(x + \delta - y) dx dy\; .
\end{equation} 
To compute $S_{k,n}(x,y)$, as before, it is convenient to
first define an auxiliary quantity 
${Q}_{k,n}(x,\Delta)$ denoting the probability that a random walk of $n$ 
steps, starting from $x_0 = x$, has $k$ 
points in the interval $(-\infty, -\Delta]$ (with $k \geq 1$) and $n-k$ points 
on the positive side, hence with no point in the interval $[-\Delta, 0]$.
The joint distribution $S_{k,n}(x,y)$ can be expressed
in terms of $Q$ as
\begin{eqnarray}\label{SQ}
{S}_{k,n}(x,y) = 
\begin{cases}
&{Q}_{k,n}(x, y-x) \;, \; x > 0 \\ 
& 0 \;, \; x < 0 \; {\rm and} \; y > 0 \\ 
&{Q}_{n-k+1,n}(-y,y-x) \;,\; x < 0 \; {\rm and} \; y < 0 \;.
\end{cases}
\end{eqnarray}

Following similar arguments leading to Eqs. (\ref{back_fp1}), 
we derive a backward integral equation, for $n\ge 1$,
\begin{eqnarray}\label{back_fptilde1}
{Q}_{k,n}(x, \Delta) &=& \int_0^\infty {Q}_{k,n-1}(x', \Delta) f(x-x') \, dx'  \\
&+& \int_{-\infty}^0 {Q}_{n-k,n-1}(-x',\Delta) f(x-x'+\Delta) dx'  \;, \nonumber
\end{eqnarray}   
starting from ${Q}_{0,0}(x, \Delta)=1$.
As before, this integral 
equation can be reduced to a linear differential recurrence equation for the 
special case, $f(x) = \frac{1}{2 b} 
\exp{(-|x|/b)}$ and subsequently solved via the generating function
method~\cite{supp_mat}. We get
\begin{eqnarray}
&&\sum_{n=0}^\infty\sum_{k=0}^n z^k s^n {Q}_{k,n}(x, \Delta) = 
\frac{1}{1-s} \nonumber \\
&&+ {A}\left(z,s, \frac{\sqrt{2}\Delta}{\sigma}\right) 
\exp{\left(-\sqrt{2(1-s)}\, \frac{x}{\sigma}\right)} \;,
\end{eqnarray} 
where ${A}(z,s,\Delta)$ has a complicated expression~\cite{supp_mat} omitted
here for clarity. 
From this result and using (\ref{SQ}) and (\ref{gappdf}), we find~\cite{supp_mat}
that as $n\to \infty$, 
$P_{k,n}(\delta)\to p_k(\delta)$ where
\begin{eqnarray}\label{GF_pk}
\sum_{k=1}^\infty z^k {p}_{k}(\delta) = 
\frac{8z}{b} e^{-2 \frac{\delta}{b}} \frac{u(z) - v(z) 
e^{-2 \frac{\delta}{b}}}{[u(z) + v(z) e^{-2 \frac{\delta}{b}}]^3}\; ,
\label{GF_pk}  
\end{eqnarray} 
with $u(z) = \sqrt{1-z} + 1$ and $v(z) = \sqrt{1-z} - 1$. Extracting 
$p_{k}$ for all $k$ from (\ref{GF_pk}) is hard. However,
one can easily extract the asymptotic behavior for large $k$, 
by analysing the $z\to 1$ limit of (\ref{GF_pk}).
This yields, for $k \gg 1$ and $\delta$ 
fixed, $p_k(\delta) \sim k^{-\frac{3}{2}} F(\delta)$ where $F(\delta)$ decays 
exponentially for large $\delta$ and represents the cut-off function. However, 
before the distribution gets
cut-off for large $\delta$, there is
a scaling regime $\delta\sim {\bar d}(k)\sim \sigma/\sqrt{2\pi k}$, with $k$
large, where we anticipate a scaling form for the gap pdf 
\begin{eqnarray}
p_{k}(\delta)\simeq ({\sqrt{k}}/{\sigma}) 
P\left({\sqrt{k}\delta}/{\sigma}\right)\,,
\end{eqnarray}
and we expect that the scaling function $P(x)$ is independent of $k$.
Indeed, taking $k\to \infty$ and $\delta\to 0$ limit in (\ref{GF_pk}) while
keeping
the scaled variable $\sqrt{k}\delta/{\sigma}$ fixed we find the scaling
function $P(x)$ satisfies
\begin{eqnarray}\label{laplace_tf}
\int_0^\infty e^{- x \lambda} \sqrt{x} P(\sqrt{x}) \, dx = 
\left(1 + \sqrt{\lambda/2}\right)^{-3}  \;.
\end{eqnarray}
This Laplace transform (\ref{laplace_tf}) can be inverted to yield finally the 
expression given in Eq. (\ref{exact_F}). The asymptotic behaviors of $P(x)$
are given by
\begin{eqnarray}\label{asympt_behavior}
P(x) \sim 
\begin{cases}
4 \sqrt{{2}/{\pi}}\;, \; x \to 0 \\
({3}/{\sqrt{8\pi}})\, x^{-4} \;, \; x \to \infty \;,
\end{cases}
\end{eqnarray}
which thus exhibits a surprising power law tail. 

\begin{figure}[hh]
\begin{center}
\includegraphics[angle=0,width = \linewidth]{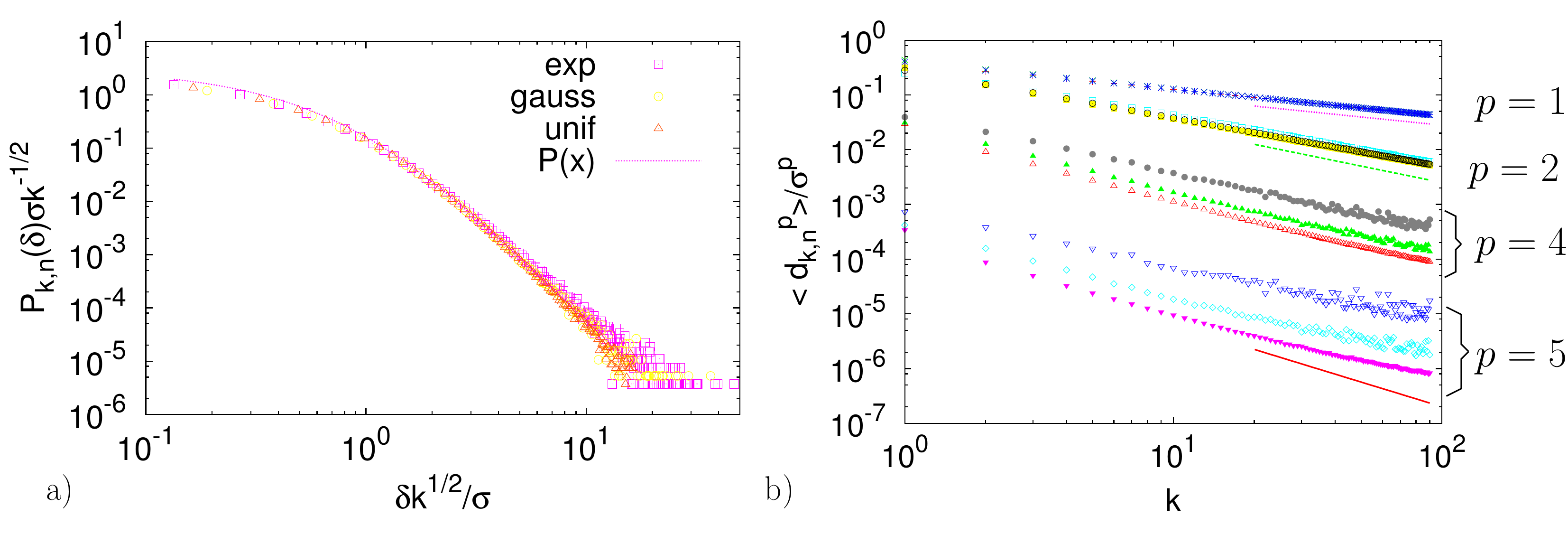}
\caption{a) : Plot of the pdf of the gaps 
$P_{k,n}(\delta) \sigma k^{-1/2}$ as a function of the 
scaling variable $\delta k^{1/2}/\sigma$ for three different 
jump distributions. The dotted line corresponds to $P(x)$ 
given Eq. (\ref{exact_F}). The good collapse of the three different 
curves indicate that $P(x)$ is universal. Note, in addition, that there 
are no fitting parameters. b) Plot of the moments 
$\langle d_{k,n}^p\rangle$ as a function of $k$ 
(for $n = 10^5$ and three different jump distributions). The data for $p=4, 5$ have been shifted downwards for clarity.}\label{fig_universal}
\end{center}
\end{figure}

The distribution $P(x)$ 
describes the typical fluctuations of $d_k$, which are of order ${\cal 
O}(k^{-1/2})$ for large $k$. Having derived it for the
special case of exponential jump distribution, it is natural
to wonder whether the same function $P(x)$ appears for
other jump distributions as well.  Remarkably, our 
numerical 
simulations show that 
$P(x)$ is indeed universal. In Fig. \ref{fig_universal} a) we 
show a plot of the (scaled) pdf of the gaps $P_{k,n}(\delta) \sigma k^{-1/2}$ 
as a function of the scaling variable $\delta k^{1/2}/\sigma$ for three 
different jump distributions:  exponential, Gaussian and uniform. The data 
shown correspond to a random walks of $n = 10^5$ and $k = 90$ and they have 
been obtained by averaging over $10^6$ independent trajectories of the random 
walk. The dotted line corresponds to $P(x)$ given in Eq. 
(\ref{exact_F}). The good collapse of these different curves, for $\delta 
k^{1/2}/\sigma \leq 1$ indicate that the typical fluctuations of $d_{k,n}$, of 
order ${\cal O}(k^{-1/2})$, are universal -- independent of $f(\eta)$ -- and 
described by $P(x)$ in Eq. (\ref{exact_F}).

In contrast to the {\em typical} fluctuations that are described by a
universal scaling function, the {\em atypically} large fluctuations
corresponding to 
$\delta \gg {\bar d}(k)\sim k^{-1/2}$ are not universal. This pdf $p_k(\delta)$
for $\delta \gg k^{-1/2}$ actually gets cut-off in a nonuniversal way, as we
have seen before for the exponential jump distribution. 
Thus there are two scales of $\delta$: a typical fluctuation which
is universal and large fluctuations that are nonuniversal.
This 
has very interesting consequences on the behavior of 
the moments $\langle d_k^p\rangle$ as a function of $k$ (for large $k$). One 
conjectures, and this is corroborated by an exact calculation for the 
exponential distribution from Eq. (\ref{GF_pk}),
\begin{eqnarray}\label{moments}
\frac{\langle d_k^p\rangle}{\sigma^p} \sim 
\begin{cases}
\frac{1}{\sqrt{2\pi}} k^{-1/2} \;, \; p =1 \\
\frac{1}{2} k^{-1} \;, \; p = 2 \\
D_3 \, (\log k) \, {k^{-3/2}} \; , \; p = 3 \\ 
D_p {k^{-3/2}} \;, \; p \geq 4 \;,
\end{cases}
\end{eqnarray}
where the amplitudes for $p<3$ are universal, while the amplitudes $D_p$ are 
not. Our numerical data, shown in Fig. \ref{fig_universal} 
b) are in agreement with these results (\ref{moments}). Indeed for $p=1$ and 
$p=2$ the value of the scaled moments $\langle d_k^p\rangle/\sigma^p$ for 
different jump distributions do coincide and exhibit a power law decay with 
$k$ in agreement with Eq. (\ref{moments}). The solid lines in Fig. 
\ref{fig_universal} indicate the power law behavior expected from Eq. 
(\ref{moments}). On the other hand, for $p=4, 5$, these scaled moments do not 
coincide and they exhibit a power law decay with, seemingly the same exponent, 
although a precise estimate of the exponent $3/2$ for higher moments is quite 
difficult.

In conclusion, we have presented exact results for the gap statistics
of symmetric random walks with a finite variance of step lengths $\sigma^2$
and found a rather rich and universal behavior independent of the
details of the jump distribution. This presents an interesting
and useful example of solvable order statistics in a correlated time series. In view of recent applications of random walks to fluctuating interfaces in $1+1$ dimensions \cite{satya_airy, schehr_airy,gyorgyi, burkhardt_nearextreme}, it will be interesting to see if the universal gap statistics found here also holds
for different boundary conditions of the interface. It would also be interesting to extend these results to cases where $\sigma$ 
is infinite such as in L\'evy flights and also to asymmetric jump
distributions.

\acknowledgments{We warmly thank Zoltan R{\'a}cz for very stimulating discussions 
at the earliest stage of this work. We thank the Lorentz Center in Leiden for 
the hospitality where part of this work was accomplished. This research was 
support by ANR grant 2011-BS04-013-01 WALKMAT and in part by the Indo-French 
Centre for the Promotion of Advanced Research under Project 4604-3.}


\begin{thebibliography}{100}



\bibitem{gumbel}
E.~J. Gumbel, {\it Statistics of Extremes}, Dover, (1958).


\bibitem{katz}
R.~W. Katz, M.~P.~Parlange and P.~Naveau, Adv. Water Resour. {\bf 25},
1287 (2002). 

\bibitem{embrecht}
P. Embrecht, C. Kl\"uppelberg, T. Mikosh, {\it Modelling Extremal
  Events for Insurance and Finance} (Springer), Berlin (1997).  

\bibitem{bouchaud_satya}
S.~N. Majumdar, J.-P. Bouchaud, Quantitative Finance {\bf 8}, 753 (2008).


\bibitem{bm97}
J.-P. Bouchaud, M. M\'ezard, J. Phys. A {\bf 30}, 7997 (1997).


\bibitem{dean_majumdar}
D.~S. Dean, S.~N. Majumdar, Phys. Rev. E {\bf 64}, 046121 (2001).

\bibitem{pldmonthus}
P. Le Doussal and C. Monthus, Physica A {\bf 317}, 140 (2003). 

\bibitem{pld_carpentier}
D. Carpentier, P. Le Doussal, Phys.Rev. E {\bf 63}, 026110 (2001);
Erratum-ibid. {\bf 73}, 019910 (2006); Y.V. Fyodorov and
J.-P. Bouchaud 2008 {\it J. Phys. A: Math. Theor.} {\bf 
41} 372001; Y.~V. Fyodorov, P. Le Doussal and A. Rosso,
J. Stat. Mech. P10005 (2009).  


%\bibitem{halpin_dprm}
%T. Halpin-Healy, Y.C. Zhang, Phys. Rep. {\bf 254}, 215 (1995).

\bibitem{satya_airy}
S.~N.~Majumdar, A.~Comtet, Phys. Rev. Lett. {\bf 92}, 225501 (2004);
J. Stat. Phys. {\bf 119}, 777 (2005).

\bibitem{schehr_airy}
G. Schehr, S.N. Majumdar, Phys. Rev. E {\bf 73}, 056103 (2006).


\bibitem{gyorgyi}
G. Gy\"orgyi, N. Moloney, G. Ozog{\'a}ny, Z. R{\'a}cz, Phys. Rev. E {\bf 75}, 021123 (2007). 


\bibitem{satya_yor}
S.~N.~Majumdar, J.~Randon-Furling, M.~J.~Kearney, M.~Yor, J. Phys. A
Math. Theor. {\bf 41}, 365005 (2008).  



\bibitem{comtet_precise}
A. Comtet, S.~N. Majumdar,  J. Stat. Mech. Theor. Exp. {\bf 06}, P06013 (2005).

\bibitem{schehr_rsrg}
G. Schehr, P. Le Doussal, J. Stat. Mech. P01009 (2010). 


\bibitem{Tracy94-96}
C.~A.~Tracy, H.~Widom, Comm. Math. Phys. {\bf 159}, 151 (1994); {\it
  ibid.} {\bf 177}, 727 (1996). 

\bibitem{sanjib_nearextreme}
S. Sabhapandit, S. N. Majumdar, Phys. Rev. Lett. {\bf 98}, 140201 (2007).

\bibitem{burkhardt_nearextreme}
T. W. Burkhardt, G. Gy{\"o}rgyi, N. R. Moloney, Z. R{\'a}cz, Phys. Rev. E {\bf 76}, 041119 (2007).


\bibitem{order_book}
H. A. David, H. N. Nagaraja, {\it Order Statistics} (third ed.), Wiley, New Jersey (2003).



\bibitem{derrida_bbm}
E. Brunet, B. Derrida, Europhys. Lett. {\bf 87}, 60010 (2009); J. Stat. Phys. {\bf 143}, 420 (2011).

\bibitem{racz_order}
N.~R.~Moloney, K.~Ozog{\'a}ny, Z. R{\'a}cz, preprint arXiv:1109.5360.   

\bibitem{tremaine_cosmo}
S. Tremaine, D. Richstone, ApJ {\bf 212}, 311 (1977). 

\bibitem{feller_book}
W. Feller, {\it An introduction to Probability Theory and its Applications}, (New York, Wiley, 1968).

\bibitem{spitzer_book}
F. Spitzer, {\it Principles of Random Walks}, Van Nostrand (Princeton), New-York (1964). 

\bibitem{queuing}
S. Asmussen, {\it Applied Probability and Queues}, (New York, Springer 2003); M. J. Kearney, J. Phys. A {\bf 37}, 8421 (2004).

\bibitem{finance}
R. J. Williams, {\it Introduction to the Mathematics of Finance} (AMS, 2006); M. Yor, {\it Exponential Functionals of Brownian Motion and Related Topics} (Berlin, Springer, 2000). 



\bibitem{pollaczek}
F. Pollaczek, C. R. Acad. Sci. Paris, {\bf 234}, 2334 (1952); J. Appl. Probab. {\bf 12}(2), 390 (1975).

\bibitem{wendel}
J. G. Wendel, Ann. Math. Statist. {\bf 31}, 1034 (1960). 


\bibitem{spitzer}
F. Spitzer, Trans. Am. Math. Soc. {\bf 82}, 323 (1956). 

\bibitem{supp_mat}
G. Schehr, S. N. Majumdar, see supplemental material.



\end{thebibliography}
\end{document}